\documentclass[aps,tightenlines,noshowpacs,preprintnumbers,prd,amsmath,letterpape,amssymb,twocolumn,osajnl,reprint]{revtex4-1}
\usepackage{amsmath}
\usepackage{graphicx}
\usepackage{sidecap}
\usepackage{caption}
\usepackage{subfigure}
\usepackage[bookmarksnumbered=true]{hyperref}
\usepackage{pythonhighlight}
\usepackage{listings}
\usepackage{xcolor}

\hypersetup{
	colorlinks = true,
	linkcolor = blue,
	anchorcolor = blue,
	citecolor = blue,
	filecolor = blue,
	urlcolor = blue
}

\definecolor{codegreen}{rgb}{0,0.6,0}
\definecolor{codegray}{rgb}{0.5,0.5,0.5} \definecolor{codepurple}{rgb}{0.58,0,0.82} \definecolor{backcolour}{rgb}{0.95,0.95,0.92}
\lstdefinestyle{mystyle}{backgroundcolor=
	\color{backcolour}, commentstyle=
	\color{codegreen}, keywordstyle=
	\color{magenta}, numberstyle=
	\tiny\color{codegray}, stringstyle=
	\color{codepurple}, basicstyle=
	\ttfamily\footnotesize, breakatwhitespace=false, breaklines=true, captionpos=b, keepspaces=true, numbers=left, numbersep=5pt, showspaces=false, showstringspaces=false, showtabs=false, tabsize=2}
\lstdefinestyle{mystyle}{ backgroundcolor=\color{backcolour}, commentstyle=\color{codegreen}, keywordstyle=\color{magenta}, numberstyle=\tiny\color{codegray}, stringstyle=\color{codepurple}, basicstyle=\ttfamily\footnotesize, breakatwhitespace=false, breaklines=true, captionpos=b, keepspaces=true, numbers=left, numbersep=5pt, showspaces=false, showstringspaces=false, showtabs=false, tabsize=2 }
\usepackage{amssymb}
\usepackage[portuges,brazilian]{babel}
\usepackage{epsfig}
\usepackage{wasysym}
\usepackage[utf8]{inputenc}
\usepackage{float}
\usepackage{booktabs}
\usepackage{mathtools}  
\usepackage{xfrac} 

\usepackage{IEEEtrantools}

\begin{document}
	
	\title{Using Mathematica software to solve ordinary differential equations and applying it to the graphical representation of trajectories\\}
	
	\author{Deyvid W. da M. Pastana$^{(a)}$\footnote{Endereço de e-mail : pastana.dwda@gmail.com} and Manuel E. Rodrigues$^{(a,b)}$\footnote{Endereço de e-mail: eleuterio@ufpa.br}}
	\affiliation{$^{(a)}$Faculdade de Ci\^{e}ncias Exatas e Tecnologia,
		Universidade Federal do Par\'{a}\\
		Campus Universit\'{a}rio de Abaetetuba, 68440-000, Abaetetuba, Par\'{a},
		Brazil\\
		$^{(b)}$Faculdade de F\'{\i}sica, Programa de P\'os-Gradua\c{c}\~ao em
		F\'isica, Universidade Federal do
		Par\'{a}, 66075-110, Bel\'{e}m, Par\'{a}, Brazil}
	
	\begin{abstract}
	We discuss the great importance of using mathematical software in solving problems in today's society. In particular, we show how to use Mathematica software to solve ordinary differential equations exactly and numerically. We also show how to represent these solutions graphically. We treat the particular case of a charged particle subject to an oscillating electric field in the xy plane and a constant magnetic field. We show how to construct the equations of motion, defined by the vectors position, velocity, electric and magnetic fields. We show how to solve these equations of Lorentz force, and graphically represent the possible trajectories. We end by showing how to build a video simulation for an oscillating electric field trajectory particle case.

		\keywords{oscillating fields, charged particle, trajectories}
	\end{abstract}
	\date{\today}
	
	\maketitle
	
	\section{Introduction}
	\label{sec:intro}
In a Bachelor's or Undergraduate course in Physics the student always comes across the subject of ordinary differential equations early on. This comes from the fact that Newton's second law is a second order ordinary differential equation. This law can be integrated exactly, in some common cases \cite{Kazunori1,Kazunori2,Symon}, or it may not have an exact analytic solution, so it must be integrated numerically \cite{Marion}. A well-known special case in which Newton's second law has no analytical solution is that of the simple pendulum \cite{Marion}. In this case, the differential equation is ordinary second order, but it is also nonlinear, which makes it impossible to integrate. A powerful method for solving ordinary differential equations that have no analytic solution is the numerical integration method. This method solves the given equation and provides a valid solution for a given initial condition and range of the independent variable. We can then obtain other functions, or quantities such as velocity and acceleration, from this numerical solution. We can also interpret the given system physically.
\par 
In physics and mathematics there has always been a need to formulate new methods for solving new problems. But there was always a limitation in the question of the possibility of the number of calculations. This was noticed more recently with the attempt to solve the three-body problem. The invention of the computer to facilitate these calculations was a milestone in physics in solving non-analytical problems. Through this tool, one can solve and interpret problems that were precluded from deeper analysis. The most recent example of the incredible amount of calculations done to obtain a result is the case of the sum of the cubes of three numbers \cite{Sum of three cubes for 42,Sum of three cubes for 42 2}. In this case, the problem is formulated as follows: Are there three real numbers raised to the cube that when added together result in the number $42$? This may seem like an easy problem, but it is a long way from being solved manually. Only with the help of a supercomputer, where you can make or check an absurd amount of calculations using about four hundred thousand home computers, can the problem be solved. The answer found was $42 = (-80538738812075974)^3 + (80435758145817515)^3 + (12602123297335631)^3$. 
\par 
The use of the computer \cite{Paulo,Julia}, with its software, to help students in elementary school \cite{Ramioro,Ramioro1} and higher education \cite{Superior,Superior 2}, is growing every year, especially in the area of exact sciences. This tool is already proven effective in solving problems and facilitating student learning in various topics in Physics \cite{Lado superior}. Computer simulation also aids student learning \cite{Lado superior 2,Lado superior 3,Lado superior 4}.
\par 
Our main goal for this article is to introduce basic codes in Mathematica software for solving ordinary differential equations exactly or numerically, and graphically representing some particle trajectories that would be impossible to do manually. It is our intention that after reading this article carefully, the physics student will be able to solve analytical and numerical problems, as well as graph their results and interpret them.
\par
The software is widely used to solve differential equations that govern the motion of charged particles subjected to electric and magnetic fields and to generate simulations of the motion for analysis, as shown in the paper \cite{D. R.}, where the authors found different confinement trajectories for the uranium isotope subjected to an oscillating magnetic field. So, software is a useful tool for the investigation of magnetic traps.
\par
The paper is presented as follows: in the section \ref{sec:2} we give an introduction to the essential commands of Mathematica software, for analytic and numerical solving of ordinary differential equations. This is done through examples such as the simple harmonic oscillator, simple pendulum, and motion with velocity proportional resistance, and also shows how to create a simulation movie. In the \ref{sec:3} section we make full use of Mathematica software, for the problem of oscillating electric and magnetic fields, showing some trajectories that are impossible to trace manually, and the construction of a movie to simulate these trajectories. In the \ref{sec:4} section we make our final considerations.

		
\section{Mathematica as a tool for solving differential equations}
	\label{sec:2}
In this section we will learn how to use Mathematica software to solve ordinary differential equations exactly and numerically. We will also show its usefulness in graphing. We will use the $12.1$ version of Mathematica software, but the commands and code can be used in any previous or newer version. 	
\par 
One of the first and most important equations in physics is that of the simple harmonic oscillator\begin{eqnarray}
\ddot{x}(t)+\omega_0^2 x(t)=0\,.\label{eqohs}
\end{eqnarray}

This equation has a unique general solution for a given initial condition, for example $x(0)=x_0$ e $\dot{x}(0)=v_0$. We can integrate this equation to find an exact solution by Mathematica software, using the following input command
\begin{lstlisting}[frame=single,mathescape=true,numbers=none,language=c++]
DSolve[D[D[x[t],t],t]+$\omega$0^2 x[t]==0,x[t],t]
\end{lstlisting}
The student should type these commands into a Mathematica notebook, hold down Shift and press Enter. The output is given by 			
\begin{lstlisting}[frame=single,mathescape=true,numbers=none,language=c++]
{{x[t]->$c_1$Cos[t$\omega$0]+$c_2$Sin[t$\omega$0]}}
\end{lstlisting}
We can add the initial conditions as follows
\begin{lstlisting}[frame=single,mathescape=true,numbers=none,language=c++]
DSolve[{D[D[x[t],t],t]+$\omega$0^2 x[t]==0,x[0]==0,x$'$[0]==1},x[t],t]
\end{lstlisting}

\begin{figure}[H]
		\begin{center}
			\includegraphics[width=0.45\textwidth]{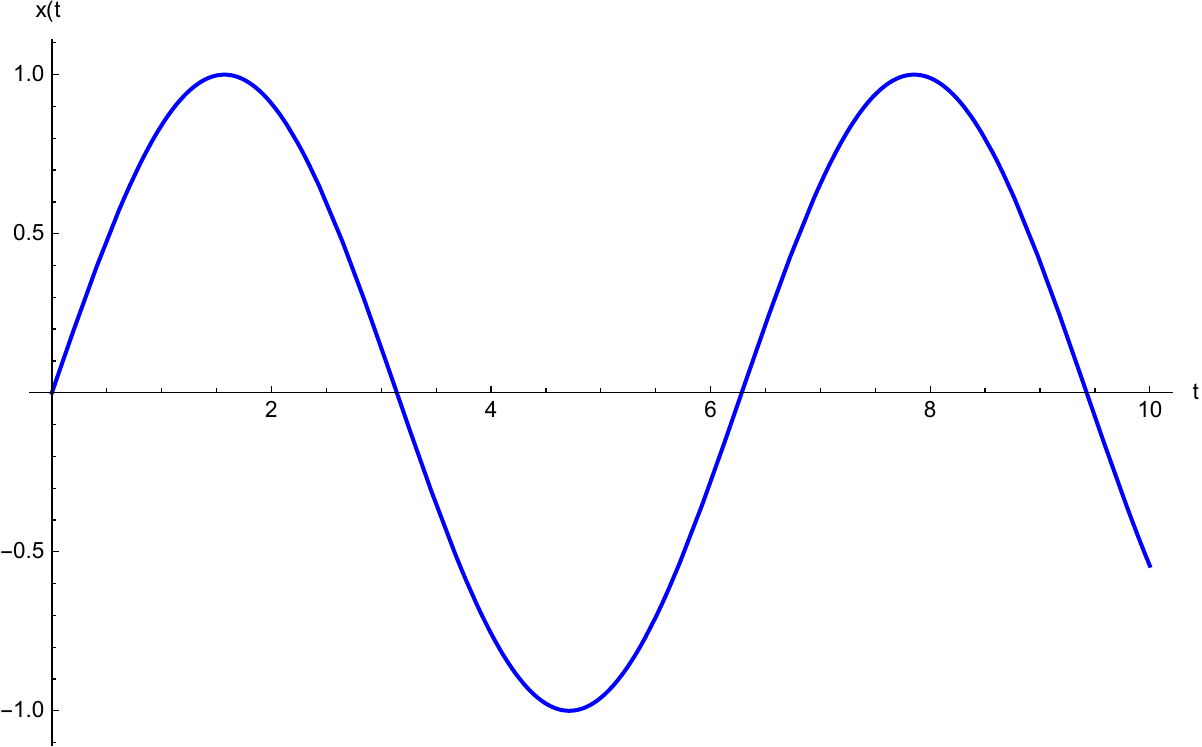}
			\begin{minipage}{8cm}
				\caption{Graph of the exact solution $x(t)$.}
				\label{fig1}
				\hspace{0.5cm}
			\end{minipage}
		\end{center}
\end{figure}

For us to be able to use this analytical solution, we must match it to some name, so we rewrite 	
\begin{lstlisting}[frame=single,mathescape=true,numbers=none,language=c++]
exactsolution=DSolve[{D[D[xe[t],t],t]+$\omega$0^2 xe[t]==0,xe[0]==0,xe$'$[0]==1},xe[t],t]
\end{lstlisting}
where $xe[t]$ is the exact solution of the equation. We can graph this exact solution with the command	
\begin{lstlisting}[frame=single,mathescape=true,numbers=none,language=c++]
Plot[{xe[t]/.exactsolution/.$\omega$0->1},{t,0,10}, PlotStyle -> Blue, AxesLabel -> {$"$t$"$, $"$x(t)$"$}]
\end{lstlisting}
The output image will be that of the figure \ref{fig1}. To call the exact solution we use the command $/.solucaoexata$, and to assign a numerical value to $\omega_0$, we use the command $/.\omega 0->1$. This same procedure can be used for any ordinary differential equation that has an exact analytic solution. The student can exercise through a question from a mathematical methods book, such as Professor Bassalo's, Arfken's, or Butkov's.

\begin{figure}[H]
		\begin{center}
			\includegraphics[width=0.45\textwidth]{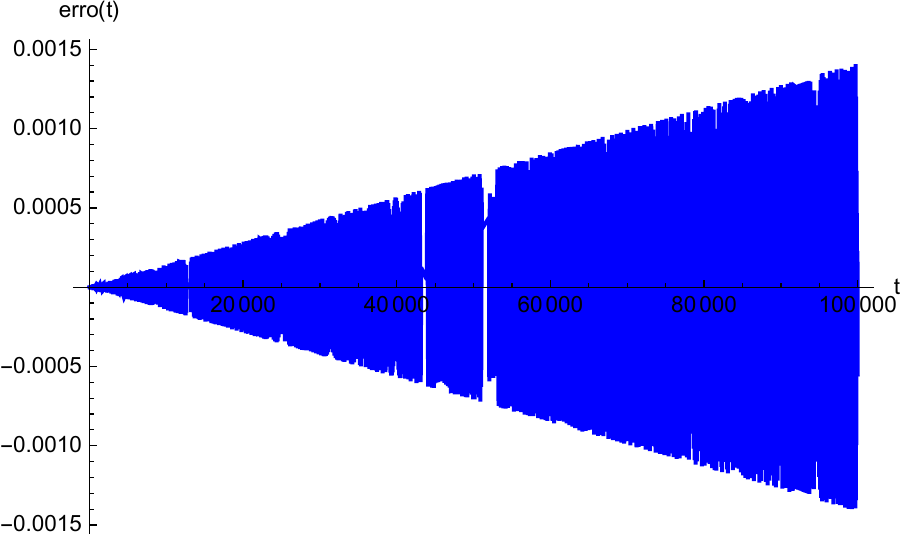}
			\begin{minipage}{8cm}
				\caption{Error function graph.}
				\label{fig2}
				\hspace{0.5cm}
			\end{minipage}
		\end{center}
\end{figure}

Now we can solve this same differential equation \eqref{eqohs}, using a numerical method. To do so, we use the following command
\begin{lstlisting}[frame=single,mathescape=true,numbers=none,language=c++]
m=1;k=1;
numericalsolution=NDSolve[{m*D[D[x[t],t],t]+k*x[t]==0,x[0]==0,x$'$[0]==1},{x[t]},{t,0,100000}]
\end{lstlisting}
We can represent graphically with the command	
\begin{lstlisting}[frame=single,mathescape=true,numbers=none,language=c++]
Plot[Evaluate[x[t] /. numericalsolution], {t, 0, 10}, PlotStyle -> Blue,AxesLabel -> {$"$t$"$, $"$x(t)$"$}]	
\end{lstlisting}	
The output is the same image as the figure \ref{fig1}, because the numerical solution has a small difference from the exact one. This can be verified as follows. Let's define a new error function, which is the subtraction of the exact solution from the numerical solution	
\begin{lstlisting}[frame=single,mathescape=true,numbers=none,language=c++]
error[t_] := 
 Simplify[Evaluate[xe[t] /. exactsolution /. $\omega$0 -> 1] - Evaluate[x[t] /. numericalsolution]]	
\end{lstlisting}	
Agora podemos plotar com	
\begin{lstlisting}[frame=single,mathescape=true,numbers=none,language=c++]
Plot[{error[t]}, {t, 0, 100000}, PlotStyle -> Blue, AxesLabel -> {$"$t$"$, $"$error(t)$"$}]
\end{lstlisting}	
The output is represented in figure \ref{fig2}. We can see that the error function is oscillatory increasing. Up to time $t=10^5$, the error function is bounded in the interval $(0.15\times 10^{-4})$. It can then be seen that the longer the time interval, the numerical solution moves further away from the exact one, and the error function grows. This can be mitigated by improving the numerical method of solving the differential equation.

\begin{figure}[H]
		\begin{center}
			\includegraphics[width=0.45\textwidth]{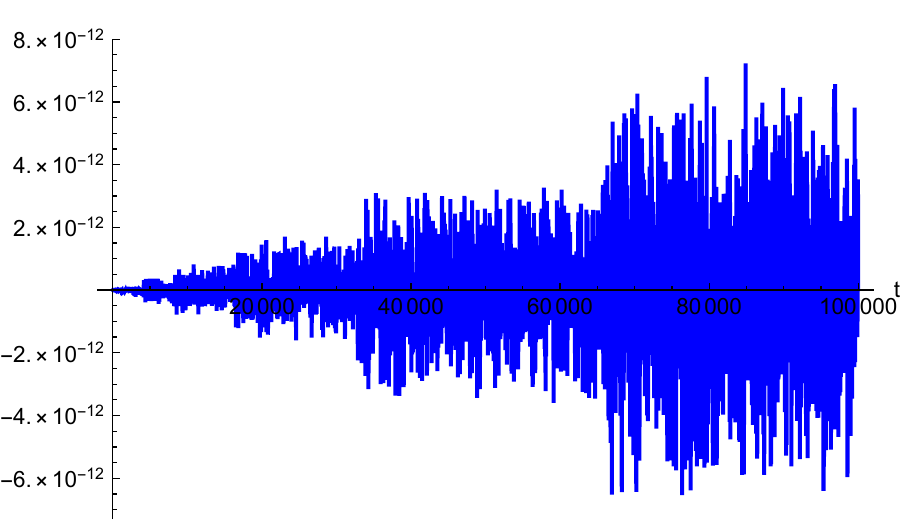}
			\begin{minipage}{8cm}
				\caption{Graph of the error$3$ function.}
				\label{fig3}
				\hspace{0.5cm}
			\end{minipage}
		\end{center}
\end{figure}

We will now improve the numerical method, using the command	
\begin{lstlisting}[frame=single,mathescape=true,numbers=none,language=c++]
numericalsolution2 = NDSolve[{m*x$''$[t] + k*x[t] == 0, x[0] == 0, x$'$[0] == 1}, {x[t]}, {t,0,100000}, MaxSteps -> Infinity, AccuracyGoal -> 10, PrecisionGoal -> 10]
\end{lstlisting}	
If we define a new error function, with the numerical solution $2$, we can see that it is restricted to a new interval $(0;4\times 10^{-5})$, thus improving the accuracy of the numerical solution. We can further improve this result. We define a new numerical solution with the command
\begin{lstlisting}[frame=single,mathescape=true,numbers=none,language=c++]
numericalsolution3 = NDSolve[{m*x$''$[t] + k*x[t] == 0, x[0] == 0, x$'$[0] == 1}, {x[t]}, {t,0,100000}, MaxSteps -> Infinity, AccuracyGoal -> 20, PrecisionGoal -> 20, WorkingPrecision -> 30]
\end{lstlisting}	
We should point out here that the computational cost, that is, the time spent by the computer, to obtain the numerical solution with this accuracy, is incredibly higher. Thus, defining the error function and its plot by 
\begin{lstlisting}[frame=single,mathescape=true,numbers=none,language=c++]
error3[t_]:= Simplify[Evaluate[xe[t] /. exactsolution /. $\omega$0 -> 1] - Evaluate[x[t] /. numericalsolution3]]	
Plot[{erro3[t]}, {t, 0, 100000}, PlotStyle -> Blue, AxesLabel -> {$"$t$"$, $"$erro(t)$"$}]
\end{lstlisting}	
The output is represented in figure \ref{fig3}. We can see that the error$3$ function is oscillatory. Up to time $t=10^5$, the error$3$ function is bounded in the interval $(0;8\times 10^{-12})$. This shows that this numerical method is very reliable, with an error of the order of $10^{-12}$.

Now we can calculate the velocity from the exact solution
\begin{lstlisting}[frame=single,mathescape=true,numbers=none,language=c++]
velocity[t_] := Evaluate[D[xe[t] /. exactsolution, t]]
\end{lstlisting}	
We can plot with the command
\begin{lstlisting}[frame=single,mathescape=true,numbers=none,language=c++]
Plot[{velocity[t] /. \[Omega]0 -> 1}, {t, 0, 50},PlotStyle -> Blue, AxesLabel -> {$"$t$"$, $"$v(t)$"$}]
\end{lstlisting}	
We could use the numerical$3$ solution, like this
\begin{lstlisting}[frame=single,mathescape=true,numbers=none,language=c++]
velocityN[t_] := Evaluate[D[x[t] /. numericalsolution, t]]
\end{lstlisting}	
By defining an error function for the velocity, we can see that the derivatives, or velocities, differ by the same order of magnitude as the original functions $xe(t)$ and $x(t)$.
\par
\begin{figure}[H]
		\begin{center}
			\includegraphics[width=0.45\textwidth]{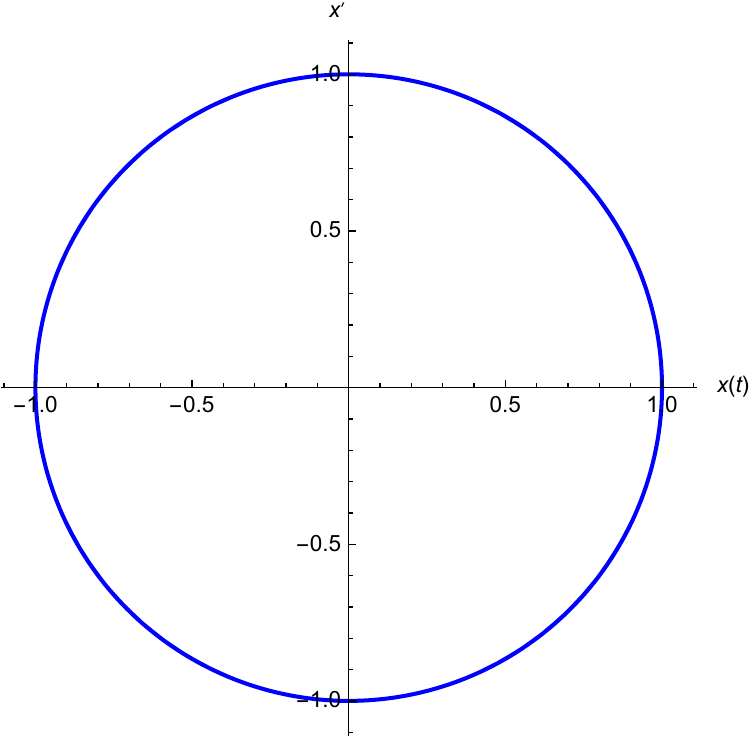}
			\begin{minipage}{8cm}
				\caption{Parametric plot of $x'\times x$.}
				\label{fig4}
				\hspace{0.5cm}
			\end{minipage}
		\end{center}
\end{figure}

 
One important tool is to make a parametric plot. We know that the position and velocity of a simple harmonic oscillator depend explicitly on time. So we can make a parametric plot $x'\times x$. For example, if $xe(t)=A\cos(\omega_0 t),ve(t)=-A\omega_0 \sin(\omega_0 t)$, then by the fundamental theorem of trigonometry $\sin^2t+\cos^2t=1$, we have $[ve/(A\omega_0)]^2+[xe/A]^2=1$, which is the equation of a circle. So doing the commands
\begin{lstlisting}[frame=single,mathescape=true,numbers=none,language=c++]
ParametricPlot[{xe[t], velocity[t]} /. exactsolution /. $\omega$0 -> 1, {t, 0, 50}, AxesOrigin -> {0, 0}, AxesLabel -> {x[t], x$'$[t]},  PlotRange -> All, ImageSize -> 300, PlotStyle -> Blue]
\end{lstlisting}	
we have the figure \ref{fig4}.

\begin{figure}[H]
		\begin{center}
			\includegraphics[width=0.12\textwidth]{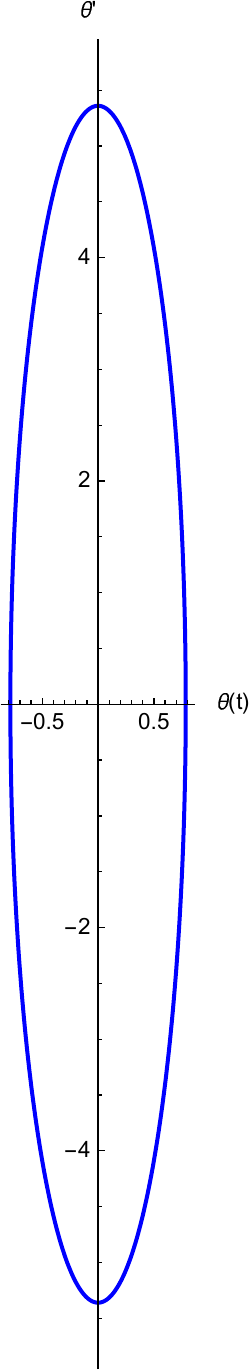}
			\begin{minipage}{8cm}
				\caption{Parametric plot of $\theta'\times \theta$.}
				\label{fig5}
				\hspace{0.5cm}
			\end{minipage}
		\end{center}
\end{figure}

Now as a practical and simple example, we can present the equation of a simple pendulum
\begin{eqnarray}
\ddot{\theta}(t)+\frac{g}{L}\sin\left[\theta(t)\right]=0\,,\label{eqps}
\end{eqnarray}
where $g$ is the acceleration of gravity, $L$ is the length of the pendulum's string and $\theta(t)$ is the angle the string makes with the vertical. As said before, the equation \eqref{eqps} is a nonlinear second order ordinary differential equation, which makes it impossible to integrate analytically. We can then use the numerical integration method presented above. Using the commands
\begin{lstlisting}[frame=single,mathescape=true,numbers=none,language=c++]
g = 9.81; L = 0.2;
numericalsolution = NDSolve[{$\theta$$''$[t] + (g/L)*Sin[$\theta$[t]] == 0, $\theta$[0] == $\pi$/4, $\theta$$'$[0] == 0}, {$\theta$[t]}, {t,0,10000}]
\end{lstlisting}	
we get the numerical solution for the simple pendulum. We can calculate the velocity of the pendulum by writing
\begin{lstlisting}[frame=single,mathescape=true,numbers=none,language=c++]
v[t_] := Evaluate[D[$\theta$[t] /. numericalsolution, t]]
\end{lstlisting}	
then we represent the parametric graph by  
\begin{lstlisting}[frame=single,mathescape=true,numbers=none,language=c++]
ParametricPlot[{$\theta$[t], v[t]} /. numericalsolution, {t, 0, 1}, AxesOrigin -> {0, 0}, AxesLabel -> {$"$$\theta$(t)$"$, $"$$\theta$$'$(t)$"$}, PlotRange -> All, ImageSize -> 100, PlotStyle -> Blue]
\end{lstlisting}	
resulting in the figure \ref{fig5}.

Finally, we will create a movie with the trajectory graphs of a particle. We will take uniform circular motion, so $x(t)=A\cos(\omega t)$ and $y(t)=A\sin(\omega t)$. Choosing $A=\omega=1$, we can create a movie with the following commands
\begin{lstlisting}[frame=single,mathescape=true,numbers=none,language=c++]
x[t_] := A*Cos[$\omega$*t]; y[t_] := A*Sin[$\omega$*t]; A = 1;$\omega$ = 1; tRange = Range[0.001, 20*Pi, 20*Pi/500];

For[k = 1, k <= Length[tRange], k++, tk = tRange[[k]]; xk = x[tk]; yk = y[tk];
  movieVector[k] = 
  Show[Graphics[{AbsolutePointSize[15], Green, Point[{xk, yk}]}, 
    PlotRange -> {{-1.1, 1.1}, {-1.1, 1.1}}], 
   ParametricPlot[{x[t], y[t]}, {t, 0, tk}, PlotStyle -> Red, 
    PlotRange -> {{-1.1, 1.1}, {-1.1, 1.1}}], AxesLabel -> {$"$x$"$, $"$y$"$},
    PlotLabel -> StringJoin[$"$t=$"$, ToString[N[tk]]]]]
    
M = Table[movieVector[k], {k, 1, 500}];

fileName = $"$FilmeMathematica.avi$"$; Export[fileName, M]

Directory[]
\end{lstlisting}	
The last command, $Directory[]$, will show you where the file is located on your computer. The generated file is in avi format. We submit the result on Youtube at the following electronic address: https://youtu.be/MLv10LFUhjA. 

Now we can move on to the next section, where we will show you some more Mathematica software tools and some trajectories where it would be impossible to represent them manually.

		
	\section{Mathematica applied to classical electromagnetism}
	\label{sec:3}

	Electromagnetism is a subject that underlies the theory of electric and magnetic fields, which are generated by sources and currents as stated in Maxwell's equations,	
	\begin{subequations}
		\label{eq:equações de Maxwell}
		\begin{align}
			\label{eq: div E}
			\nabla\cdot\vec{E}&=\frac{\rho}{\epsilon_{0}}\\
			\label{eq: div B}
			\nabla\cdot\vec{B}&=0\\
			\label{eq: rot E}
			\nabla\times\vec{E}&=-\frac{\partial \vec{B}}{\partial t}\\
			\label{eq: rot B}
			\nabla\times\vec{B}&=\mu_{0}\vec{J}+\mu_{0}\epsilon_{0}\frac{\partial\vec{E}}{\partial t}
		\end{align}
	\end{subequations}
	
	Once a system is established, the conservation of charges locally is represented by the continuity equation,
	\begin{equation}\label{eq: continuidade}
		\nabla\cdot\vec{J}+\frac{\partial\rho}{\partial t}=0.
	\end{equation}
	
	Therefore, Maxwell's equations govern the form fields can take such that if an electric field is reproduced experimentally, it will in turn produce a magnetic field corresponding with Maxwell's equations, and the same can be said for current density and charge desity by means of the continuity equation.
	
	Given a magnetic field and an electric field, we can check whether they satisfy Maxwell's equations and the continuity equation easily using Mathematica software by using the codes
		\begin{lstlisting}[frame=single,mathescape=true,numbers=none,language=c++]
		
		Div[E,{x, y, z}] == $ \rho/\epsilon_{0} $
		
		Div[B,{x, y, z}] == 0
		
		Curl[E,{x, y, z}] == -D[B,t]
		
		Curl[B,{x ,y ,z }] == $ \mu_{0} $J+$\mu_{0}\epsilon_{0}$D[E,t]
		
		Div[J,{x, y, z}]+D[$\rho$,t] == 0
	\end{lstlisting}
	where the commands $ Div[]\ \text{e}\ Curl[] $ represent the divergent and rotational, respectively, of the fields.
	
	The effects of electric and magnetic fields on an electric charge $ Q $ are represented by the Lorentz force,
	\begin{equation}\label{eq: força de Lorentz}
		\vec{F}=Q\left(\vec{E}+\vec{v}\times\vec{B}\right).
	\end{equation}
	
	Equating the (\ref{eq: Lorentz force}) to Newton's second law of motion, we obtain the equations of motion for this system, with which we can analyze the dynamics of a charged particle under the action of electric and magnetic fields.
	
	\subsection{Oscillating fields and charged particles}
	\label{subsec:Campos oscilantes}
		
	In this section we will show how Mathematica software can be used to solve the equations of motion and generate simulations of possible trajectories for an electrically charged particle subjected to an oscillating electric field and a uniform magnetic field.
	
	\subsubsection{$\vec{E}$ oscillating $ (x) $ and $\vec{B}$ uniform $ (z) $}
	\label{subsubsec: 1}
	
	Consider a charged particle of mass $ M $ and charge $ Q $, subjected to the fields
	\begin{align}
		\label{eq: campo E do arquivo 0}
		\vec{E}&=E_{0}\cos(\omega t)\hat{i}\\
		\label{eq: campo B do arquivo 0}
		\vec{B}&=\frac{M\omega_{0}}{Q}\hat{k}
	\end{align}
	
	Note that the fields (\ref{eq: campo E do arquivo 0}) and (\ref{eq: campo B do arquivo 0}) do not satisfy the Ampère-Maxwell law, equation (\ref{eq: rot B}), in the absence of sources and currents. However, all of Maxwell's equations are satisfied if one validates
	\begin{equation}\label{eq:densidade de corrente 0}
		\vec{J}=\epsilon_{0}E_{0}\omega\sin(\omega t)\hat{i}.
	\end{equation} 
	
	This is in accordance with the continuity equation (\ref{eq: continuidade}), because the current density vector (\ref{eq:densidade de corrente 0}) is uniform and therefore has zero divergence.
	
	An electric field oscillating in time and uniform in space satisfies all Maxwell's equations for any constant magnetic field if $ \vec{J}=-\sfrac{\partial \vec{E}}{\partial t} $. This condition immediately implies, $\nabla\cdot\vec{J}=0$. Given that
	\begin{equation}
		i=\int_{V}\nabla\cdot\vec{J}dV,
	\end{equation}
no current located within a volume $ V $ exits by a flow through a surface $ A $, or, furthermore, an external current does not enter within the volume of consideration $ V $ through a surface $ A $. Thus, there is no variation of a net charge density as a function of spatial variables, which is characteristic of a stationary magnetic field \cite{Jackson}. Therefore, the current density (\ref{eq:densidade de corrente 0}) obeys the laws of magnetostatics, as it must, since we are considering a uniform magnetic field.
	
	Replacing the fields (\ref{eq: campo E do arquivo 0}) and (\ref{eq: campo B do arquivo 0}) in the Lorentz force, we obtain the equations of motion
	\begin{subequations}
		\label{eq: equação de movimento arq 0}
		\begin{gather}
			QE_{0}\cos(\omega t)+M(\omega_{0}\dot{y}-\ddot{x})=0\\
			-M\left(\omega_0 \dot{x}+\ddot{y}\right)=0\\ 
			-M\ddot{z}=0.
		\end{gather}
	\end{subequations}
	
	Given the initial conditions $ x(0)=y(0)=z(0)=\dot{x}(0)=\dot{y}(0)=\dot{z}(0)=0 $, we can solve the coupled ODE system (\ref{eq: equação de movimento arq 0}), obtaining the solutions
	\begin{gather}
		x(t)=\frac{E_0Q[\cos(\omega_0 t)-\cos(\omega t)]}{M(\omega^2-\omega_0^2)}\\ 
		y(t)=\frac{E_0Q\left[\omega_0 \sin(\omega t)-\omega sin(\omega_{0} t)\right]}{M(\omega^3 - \omega \omega_0^2)}\\
		z(t)=0.
	\end{gather}
	
\begin{figure}[H]
		\begin{center}
			\includegraphics[width=0.45\textwidth]{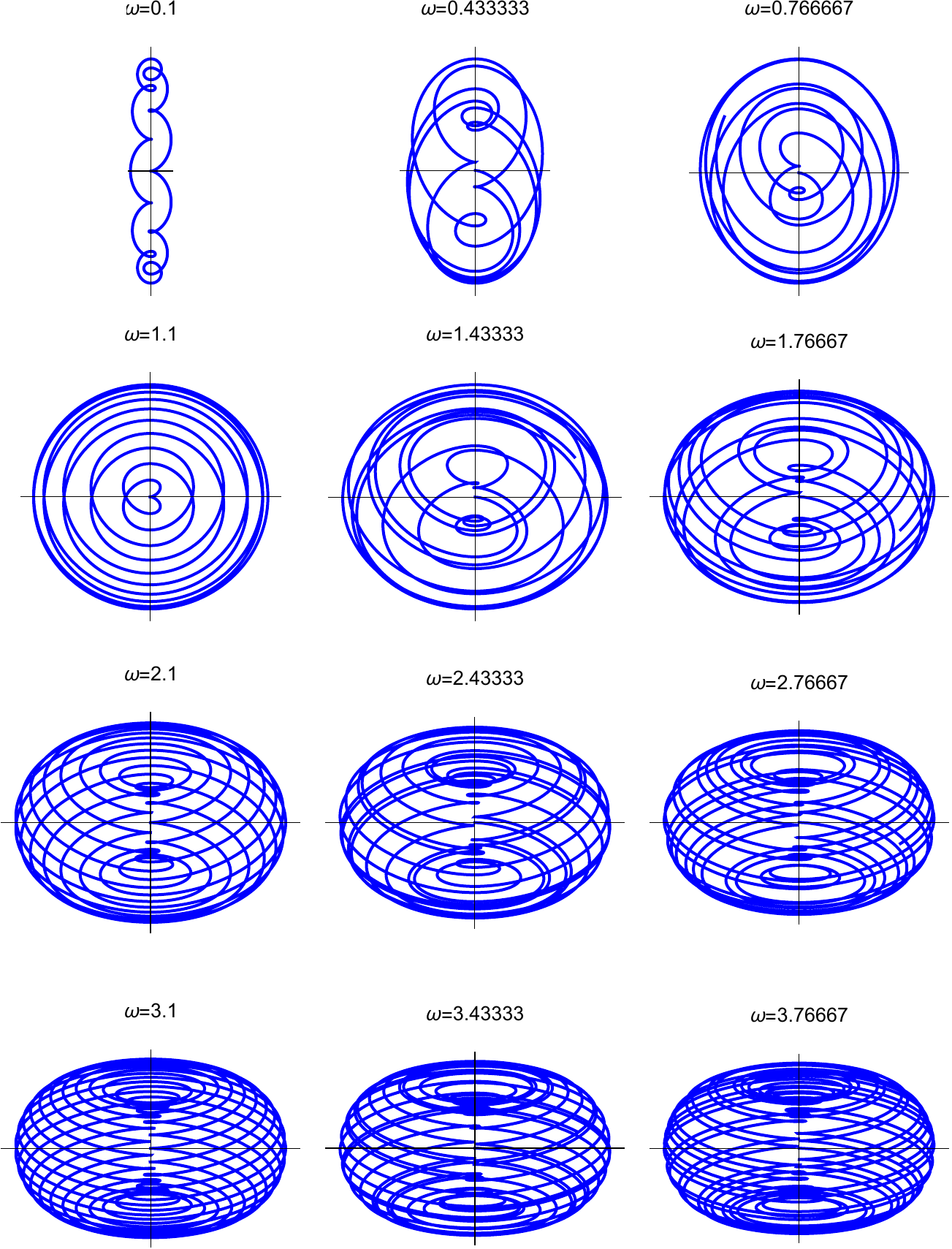}
			\begin{minipage}{8cm}
				\caption{Trajectories of a charged particle subjected to an oscillating electric field and a uniform magnetic field, with the oarameter $ M=E_0=\omega_{0}=Q=1 $ and initial conditions $ x(0)=y(0)=z(0)=\dot{x}(0)=\dot{y}(0)=\dot{z}(0)=0 $.}
				\label{fig: trajetoria arq 0}
				\hspace{0.5cm}
			\end{minipage}
		\end{center}
	\end{figure}

	Setting the parameters $ M=E_0=\omega_{0}=Q=1 $ we can express the motions of the charged particle, which occur in the $ xy $ plane, for different values of $homega$, when subjected to the fields (\ref{eq: campo E do arquivo 0}) and (\ref{eq: campo B do arquivo 0}), as shown in the figure \ref{fig: trajetoria arq 0}.
	
	Using Mathematica software, this procedure can be verified, along with obtaining graphs, from the commands
	
	\begin{lstlisting}[frame=single,mathescape=true,numbers=none]
		
		F = {Fx,Fy,Fz};r = {x[t],y[t],z[t]};EF = {Ex,Ey,Ez};BF = {Bx,By,Bz};v = D[r,t];
		
		F = Q*(EF+Cross[v,BF])
		
		eq1 = F-M*D[r,{t,2}] == 0//Thread; eq1//ColumnForm
		
		initial = {x[0] == 0,x'[0] == 0,y[0] == 0,y'[0] == 0,z[0] == 0,z'[0] == 0};
		
		fields = Thread/@{BF->{0,0,M*$\omega 0/Q$},EF->{E0*\cos[$ \omega $*t],0,0}}//Flatten
		
		eqs = Join[eq1,inicial] /.fields
		
		solution = DSolve[eqs,{x[t],y[t],z[t]},t]//Flatten//FullSimplify
		
		values = {M->1,E0->1,$ \omega $0->1,$ Q $->1}
		
		graphic[t_,$ \omega $] := {x[t],y[t],z[t]}/.solution/.values
		
		graphic[$\omega$_] := ParametricPlot[point[t,$ \omega $][[{1,2}]]//Evaluate,{t,0,20*$ \pi $},PlotLabel -> StringJoin["$\omega$=", ToString[N[$\omega$]]],Ticks->None,DisplayFunction->Identity, PlotStyle -> Blue]
		
		table := Table[graphic[$\omega$],{$\omega$,0.1,4,1/3}]
		
		Show[GraphicsArray[Partition[table,3]]]
	\end{lstlisting}
	where Table[grafico[$ \omega $],{$ \omega $,0.1,4, 1/3}] is a command that collects the points generated by the solutions, for different values $ \omega $, building a table, which is called with Show[GraphicsArray[Partition[tabela,3]]] to generate the graphs of the trajectories illustrated in the figure \ref{fig: trajetoria arq 0}.

	
	\subsubsection{\label{subsubsec: 2}$ \vec{E} $ oscillating in two directions $ (xy) $ and $ \vec{B} $ uniform in a $ (z) $}
	
	Given a particle of charge $ Q $ and mass $ M $ under the action of the fields
	
	\begin{align}
	\label{eq: campo E do arquivo 1}
	\vec{E}&=E_{0}\cos(\omega t)\hat{i}+E_{0}\sin(\omega t)\hat{j}\\
	\label{eq: campo B do arquivo 1}
	\vec{B}&=\frac{M\omega_{0}}{Q}\hat{k},
	\end{align}
	
	The equations of motion for this system will be
	\begin{subequations}
		\label{eq: equação de movimento arq 1}
		\begin{gather}
			QE_{0}\cos(\omega t)+M(\omega_{0}\dot{y}-\ddot{x})=0\\
			QE_{0}\sin(\omega t)-M\left(\omega_0 \dot{x}+\ddot{y}\right)=0\\ 
			-M\ddot{z}=0.
		\end{gather}
	\end{subequations}
	
	The System (\ref{eq: equação de movimento arq 1}), considering the initial conditions $ x(0)=y(0)=z(0)=\dot{x}(0)=\dot{y}(0)=\dot{z}(0)=0 $, has as a general solution
	\begin{gather}
		x(t)=\frac{E_0 Q \left[\omega+\omega_0-\omega_0 \cos(t\omega)-\omega\cos(t\omega_0)\right]}{(m\omega_0\omega_0(\omega_0+\omega_0))}\\ 
		y(t)=\frac{E_0 Q \left[\omega\sin(t \omega_0)-\omega_0\sin(t \omega)\right]}{m \omega \omega_0(\omega+\omega_0)}\\
		z(t)=0
	\end{gather}
	
\begin{figure}[H]
		\begin{center}
			\includegraphics[width=0.45\textwidth]{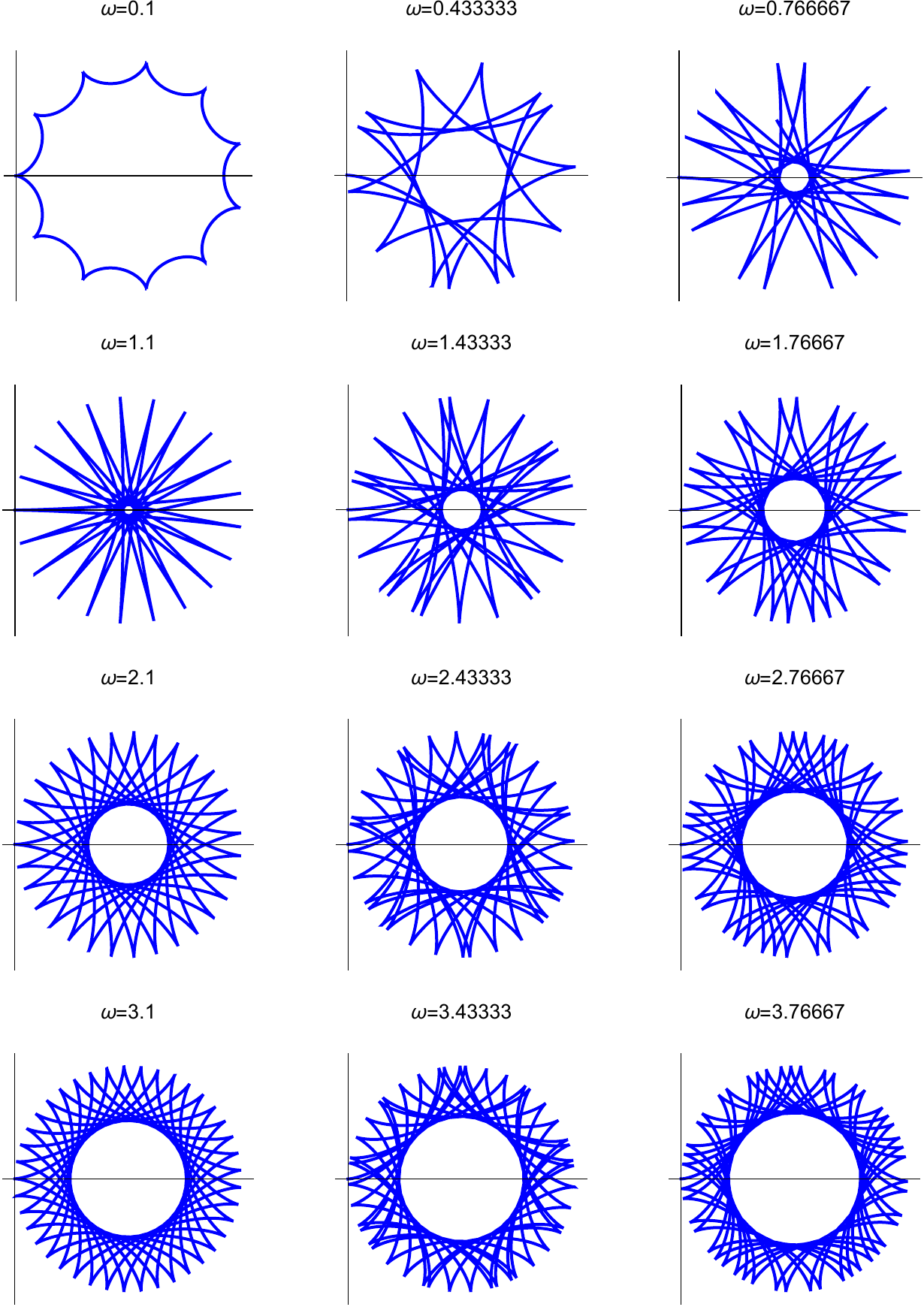}
			\begin{minipage}{8cm}
				\caption{Trajectories of a charged particle subjected to an oscillating electric field in two directions and a uniform magnetic field in one, with the parameters $ M=E_0=\omega_{0}=Q=1 $ and initial conditions $ x(0)=y(0)=z(0)=\dot{x}(0)=\dot{y}(0)=\dot{z}(0)=0 $.}
				\label{fig: trajetoria arq 1}
				\hspace{0.5cm}
			\end{minipage}
		\end{center}
	\end{figure}

	By defining the parameters $ M=E_0=\omega_{0}=Q=1 $, we obtain the trajectories that the particle describes in a plane, for different values of $\omega$, in the $ xy $ plane, as shown in the figure \ref{fig: trajetoria arq 1}.
	
	The trajectories illustrated in the figure \ref{fig: trajetoria arq 1} were generated in Mathematica from the commands
	\begin{lstlisting}[frame=single,mathescape=true,numbers=none]
		
		F = {Fx, Fy, Fz}; r = {x[t], y[t], z[t]}; EF = {Ex, Ey,Ez}; BF = {Bx, By, Bz}; v = D[r, t];
		
		F = Q*(EF + Cross[v, BF])
		
		eq1 = F - M*D[r, {t, 2}] == 0 // Thread; eq1 // ColumnForm
		
		initial = {x[0] == 0, x'[0] == 0, y[0] == 0, y'[0] == 0, z[0] == 0, z'[0] == 0};
		
		fields = Thread /@ {BF->{0, 0, m*$\omega$0/q},EF->{E0*Cos[$\omega$*t], E0*Sin[$\omega$*t], 0}} // Flatten
		
		eqs = Join[eq1, initial] /. fields
		
		solution = DSolve[eqs, {x[t], y[t], z[t]}, t] // Flatten // FullSimplify
		
		values = {M -> 1, E0 -> 1, $\omega$0 -> 1, Q -> 1}
		
		point[t_, $ \omega $_] := {x[t], y[t], z[t]} /. solution /. values
		
		graphic[$\omega$_]:=ParametricPlot[point[t,$\omega$][[{1,2}]]//Evaluate,{t,0,20*Pi},PlotLabel->StringJoin["$\omega$=",ToString[N[$\omega$]]],Ticks->None,DisplayFunction->Identity, PlotStyle -> Blue]
		
		table := Table[graphic[$ \omega $], {$ \omega $, 0.1, 4, 1/3}]
		
		Show[GraphicsArray[Partition[tabela, 3]]]
	\end{lstlisting}

	
	\subsubsection{\label{subsubsec: 3}$\vec{E}$ oscillating in two directions $ (xy) $ and $\vec{B}$ uniform $ (x) $}

	Let's consider the electric field used in the case \ref{subsubsec: 2}, this time with a constant magnetic field $ B_{x} $ in the direction of the $ x $ axis, i.e.,
	\begin{align}
		\label{eq: campo E do arquivo 2}
		\vec{E}&=E_{0}\cos(\omega t)\hat{i}+E_{0}\sin(\omega t)\hat{j}\\
		\label{eq: campo B do arquivo 2}
		\vec{B}&=B_{x}\hat{i}
	\end{align}

	A system in which a charged particle of mass $ M $ and charge $ Q $, subjected to the action of the fields (\ref{eq: campo E do arquivo 2}) and (\ref{eq: campo B do arquivo 2}), leads to the equations of motion
	\begin{subequations}
		\label{eq: equação de movimento arq 2}
		\begin{gather}
			E_{0}Q\cos(\omega t)-M\ddot{x}=0\\
			Q\left[E_{0}\sin(\omega t)+B_{x}\dot{z}\right]-M\ddot{y}=0\\
			-B_{x}Q\dot{y}-M\ddot{z}=0
		\end{gather}
	\end{subequations}

	Assuming that the initial conditions are $x(0)=y(0)=z(0)=\dot{x}(0)=\dot{y}(0)=\dot{z}(0)=0$, we obtain the following solution
	\begin{align}
		\label{eq: sol da equação de movimento arq 2 x}
		x(t)&=-\frac{E_0 Q \cos(\omega t)-1}{M\omega^2}\\
		\label{eq: sol da equação de movimento arq 2 y}
		y(t)&=\frac{E_0 M \left[-M \omega \sin(\sfrac{B_x Q t}{M}) + B_x Q \sin(\omega t)\right]}{B_x^3 Q^2 - B_x M^2 \omega^2}\\
		z(t)&=\frac{E_0 (-B_x^2 Q^2 + M^2 \omega^2-M^2 \omega^2 \cos(\sfrac{Bx Q t}{M})}{B_x^3 Q^2 \omega-B_x M^2 \omega^3}\nonumber\\
		\label{eq: sol da equação de movimento arq 2 z}
		&+\frac{E_0 (B_x^2 Q^2 \cos(\omega t))}{B_x^3 Q^2 \omega-B_x M^2 \omega^3}
	\end{align}
	
\begin{figure}[H]
		\centering
		\begin{minipage}{8cm}
			\centering
			\includegraphics[width=1\textwidth,height=8.5cm]{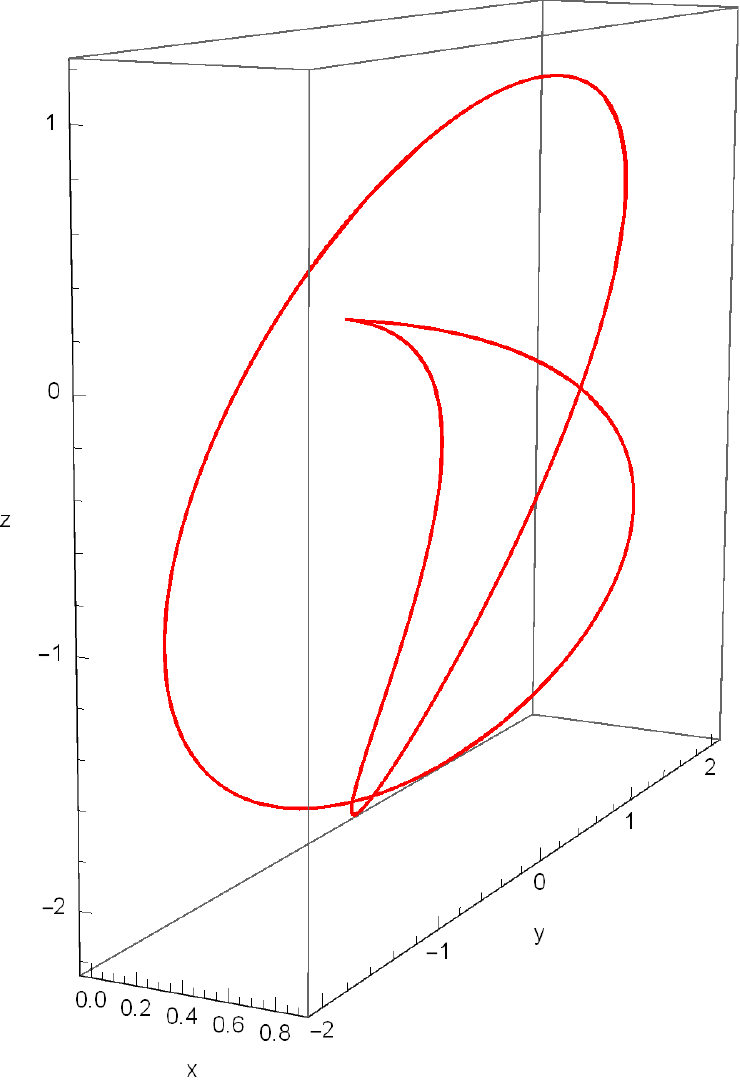}
			\captionof{figure}{Trajectory of a charged particle subjected to fields (\ref{eq: campo E do arquivo 2}) and (\ref{eq: campo B do arquivo 2}), with the parameters $ M=E_0=B_{x}=Q=1 $, $\omega=1,5$ and initial conditions $ x(0)=y(0)=z(0)=\dot{x}(0)=\dot{y}(0)=\dot{z}(0)=0 $, in a time interval [$ 0,20\pi $].}
			\label{fig: trajetoria arq 2.1}
		\end{minipage}\hspace*{2cm}
		\begin{minipage}{8cm}
			\centering
			\includegraphics[width=1\textwidth,height=8.5cm]{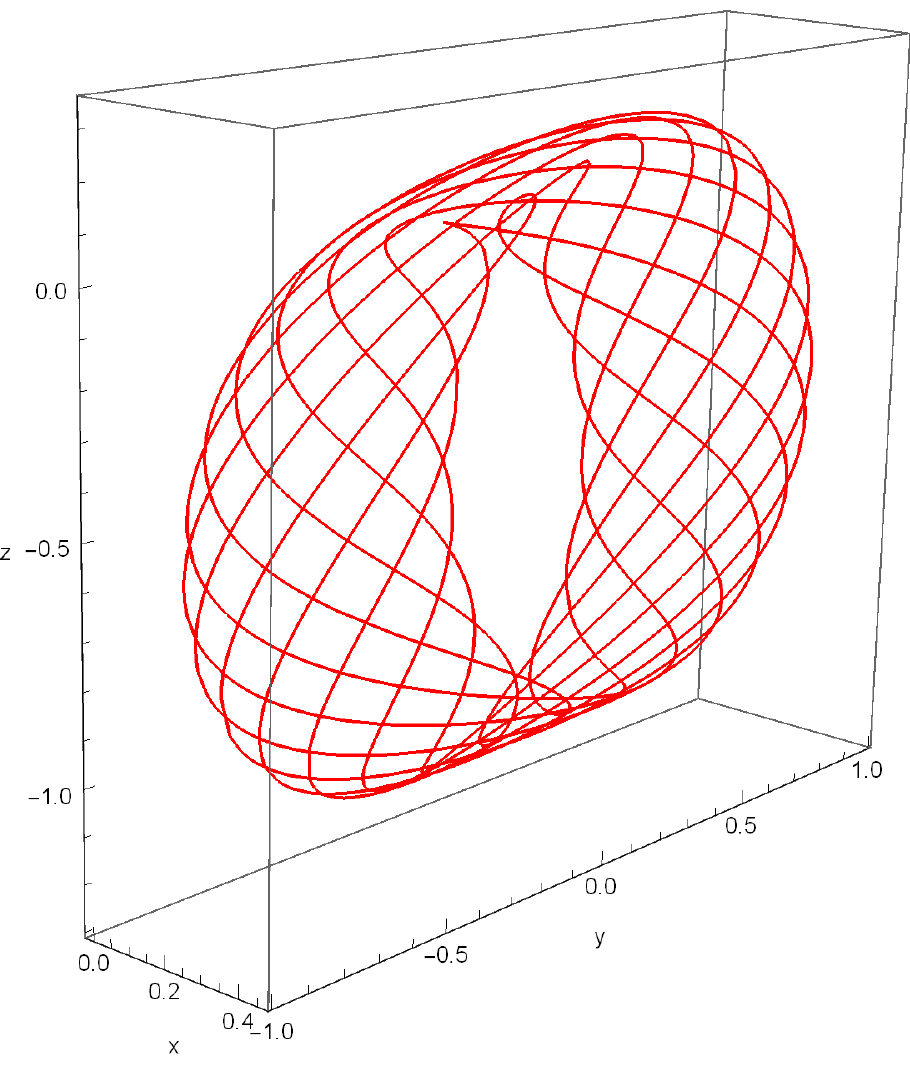}
			\captionof{figure}{Trajectory of a charged particle subjected to fields (\ref{eq: campo E do arquivo 2}) and (\ref{eq: campo B do arquivo 2}), with the parameters $ M=E_0=B_{x}=Q=1 $, $\omega=2.1$ and initial conditions $ x(0)=y(0)=z(0)=\dot{x}(0)=\dot{y}(0)=\dot{z}(0)=0 $, in a time interval [$ 0,20\pi $].}
			\label{fig: trajetoria arq 2.2}
		\end{minipage}
	\end{figure}

\begin{figure}[H]
		\centering
		\begin{minipage}{8cm}
			\centering
			\includegraphics[width=1\textwidth,height=8cm]{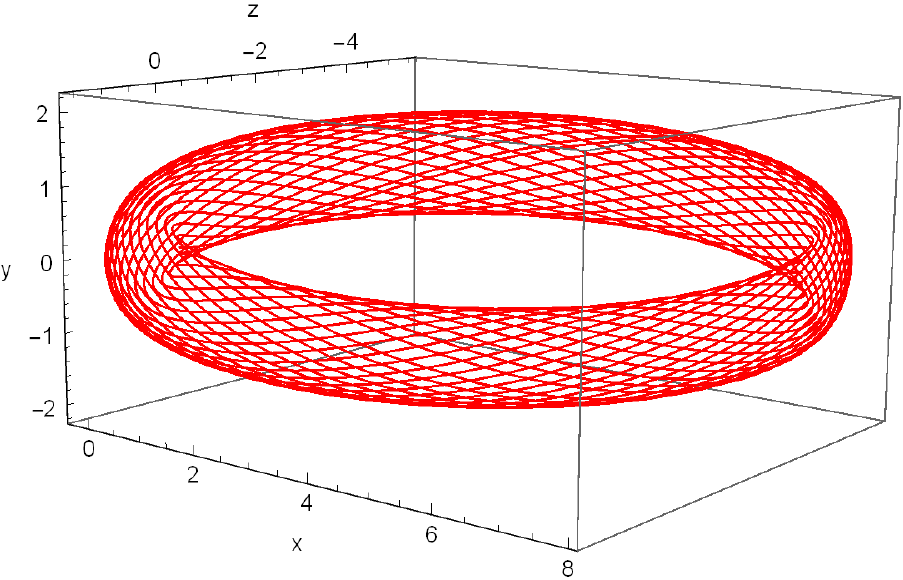}
			\captionof{figure}{Trajectory of a charged particle subjected to fields (\ref{eq: campo E do arquivo 2}) and (\ref{eq: campo B do arquivo 2}), with the parameters $ M=E_0=B_{x}=Q=1 $, $\omega=0.51$ and initial conditions $x(0)=y(0)=z(0)=\dot{x}(0)=\dot{y}(0)=\dot{z}(0)=0$, in a time interval [$ 0,100\pi $].}
			\label{fig: trajetoria arq 2.4}
		\end{minipage}
	\end{figure}

	Defining the values of the constants as $ E_0=Q=M=B_x=1 $, we get the trajectories shown in the figures \ref{fig: trajetoria arq 2.1}, \ref{fig: trajetoria arq 2.2} and \ref{fig: trajetoria arq 2.4}.
	
	\onecolumngrid
	\twocolumngrid
	
	The solutions (\ref{eq: sol da equação de movimento arq 2 x}), (\ref{eq: sol da equação de movimento arq 2 y}) and (\ref{eq: sol da equação de movimento arq 2 z}), can be obtained in Mathematica from the commands
	
	\begin{lstlisting}[frame=single,mathescape=true,numbers=none]
		
		F = {Fx, Fy, Fz}; r = {x[t], y[t], z[t]}; EF = {Ex, Ey, Ez}; BF = {Bx, By, Bz}; v = D[r, t];
		
		F = Q*(EF + Cross[v, BF])
		
		eq1 = F - M*D[r, {t, 2}] == 0 // Thread; eq1 // ColumnForm
		
		initial = {x[0] == 0, x'[0] == 0, y[0] == 0, y'[0] == 0, z[0] == 0, z'[0] == 0};
		
		fields = Thread /@ {BF -> {Bx, 0, 0}, EF -> {E0*Cos[$ \omega $*t],E0*Sin[$ \omega $*t], 0}} // Flatten
		
		eqs = Join[eq1, initial] /. fields
		
		DSolve[eqs, {x[t], y[t], z[t]}, t] // Flatten // FullSimplify
		
	\end{lstlisting}

	Once the solutions are obtained, we use ParametricPlot3D to build the computer simulation of the motion of the charged particle illustrated in figure \ref{fig: trajetoria arq 2.1}, from the codes \vspace{15pt}
	
	\begin{lstlisting}[frame=single,mathescape=true,numbers=none]
		
		values = {M -> 1, E0 -> 1, Bx -> 1, Q -> 1}
		
		eqx[t_] := -((E0 q (-1+Cos[t $\omega$]))/(m $\omega$^2)); 
		eqy[t_] := (E0 m (-m $\omega$ Sin[(Bx q t)/m] + Bx q Sin[t $\omega$]))/(Bx^3 q^2 - Bx m^2 $\omega$^2); 
		eqz[t_] := (E0 (-Bx^2 q^2 + m^2 $\omega$^2 - m^2 $\omega$^2 Cos[(Bx q t)/m] + Bx^2 q^2 Cos[t $\omega$]))/(Bx^3 q^2 $\omega$ - Bx m^2 $\omega$^3);
		
		ParametricPlot3D[{eqx[t], eqy[t], eqz[t]} /. $\omega$ -> 1.5 /. valores, {t, 0, 20*Pi}, AxesLabel -> {"x", "y", "z"}, Mesh -> None, PlotRange -> All, PlotStyle -> {Red, Directive[EdgeForm[]], Thickness[0.005],		Opacity[1]}, ImageSize -> Medium]
		
	\end{lstlisting}
	
	The figures \ref{fig: trajetoria arq 2.2}  and \ref{fig: trajetoria arq 2.4}, are easily obtained by varying the value of $\omega$ set in the ParametricPlot3D command as $\omega$ $ -\!\!> $ 1.5, for the values $ 2.1,\ \text{e}\ 0.51 $, respectively.	
	
	\subsubsection{\label{subsubsec: 4}$\vec{E}$ oscillating in two directions $ (xy) $ and $\vec{B}$ uniform in a $ (y) $}
	
	Consider a particle of mass $ M $ and charge $ Q $ subjected to the effects of an oscillating electric and magnetic field uniform in the direction of the axis $ x $,
	\begin{align}
		\label{eq: campo E do arquivo 3}
		\vec{E}&=E_{0}\cos(\omega t)\hat{i}+E_{0}\sin(\omega t)\hat{j}\\
		\label{eq: campo B do arquivo 3}
		\vec{B}&=B_{y}\hat{j}.
	\end{align}
	
	The equations of motion for the system formed by the particle and the fields (\ref{eq: campo E do arquivo 3}) and (\ref{eq: campo B do arquivo 3}) are
	\begin{subequations}
		\begin{gather}
		Q \left[E_0 \cos(\omega t)-B_y\dot{z}\right]-M\ddot{x}=0\\
		QE_0 \sin(\omega t)-M\ddot{y}=0\\
		QB_y \dot{x}-M\ddot{z}=0
		\end{gather}
	\end{subequations}
	
	Assuming that the initial conditions are given by $x(0)=y(0)=z(0)=\dot{x}(0)=\dot{y}(0)=\dot{z}(0)=0$, we get the solution
	\begin{gather}
		\label{eq: sol da equação de movimento arq 3 x}
		x(t)=\frac{QME_0\left[\cos\left(\sfrac{B_y Q t}{M}\right)-\cos(\omega t)\right]}{M^2 \omega^2-B_y^2 Q^2}\\
		\label{eq: sol da equação de movimento arq 3 y}
		y(t)=\frac{QE_0\left[\omega t -\sin(\omega t)\right]}{M\omega ^2}\\
		\label{eq: sol da equação de movimento arq 3 z}
		z(t)=\frac{QE_0\left[B_y Q\sin (\omega t)-M\omega\sin\left(\sfrac{B_yQt}{M}\right)\right]}{B_y^2 Q^2\omega-M^2\omega ^3}
	\end{gather}

\begin{figure}[H]
		\begin{center}
			\begin{minipage}{8cm}
				\centering
				\includegraphics[width=1\textwidth,height=8cm]{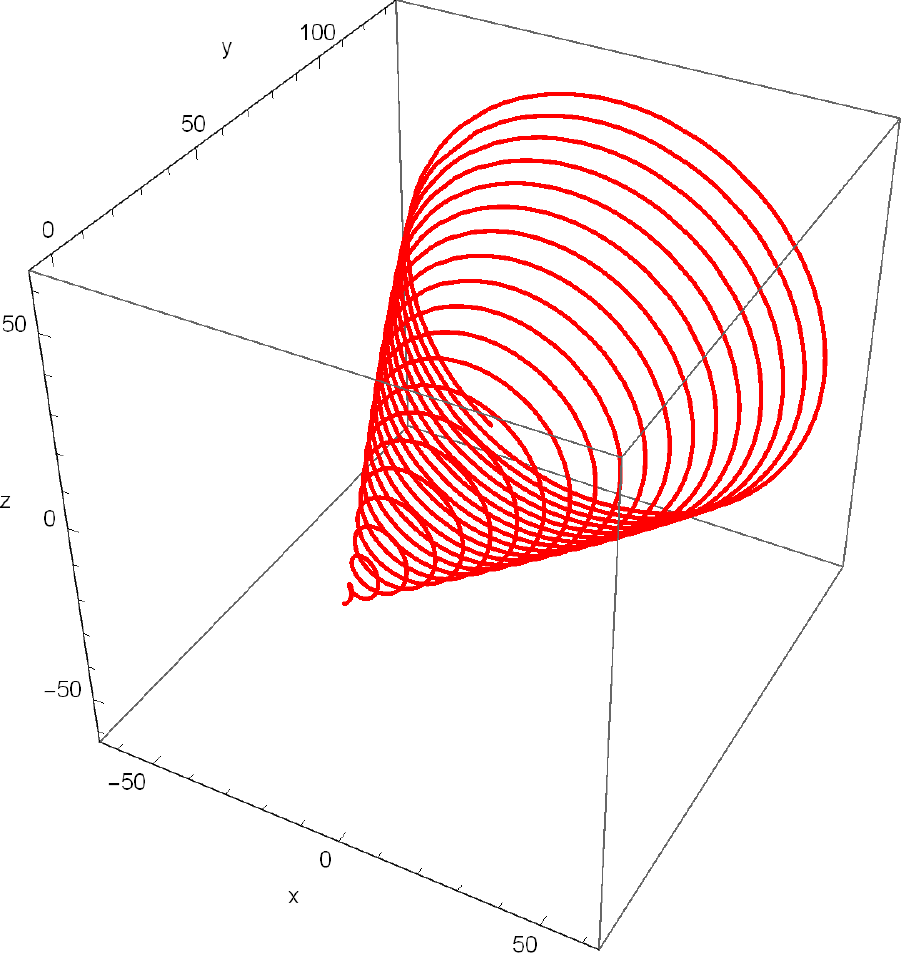}
				\captionof{figure}{Trajectory of a charged particle subjected to fields (\ref{eq: campo E do arquivo 2}) and (\ref{eq: campo B do arquivo 2}), with the parameters $ M=E_0=B_{y}=Q=1 $, $\omega=0.99$ and initial conditions $x(0)=y(0)=z(0)=\dot{x}(0)=\dot{y}(0)=\dot{z}(0)=0$, in a time interval [$ 0,40\pi $].}
				\label{fig: trajetoria arq 3.3}
			\end{minipage}
		\end{center}
	\end{figure}

\begin{figure}[H]
		\begin{center}
			\begin{minipage}{8cm}
				\centering
				\includegraphics[width=1\textwidth,height=8cm]{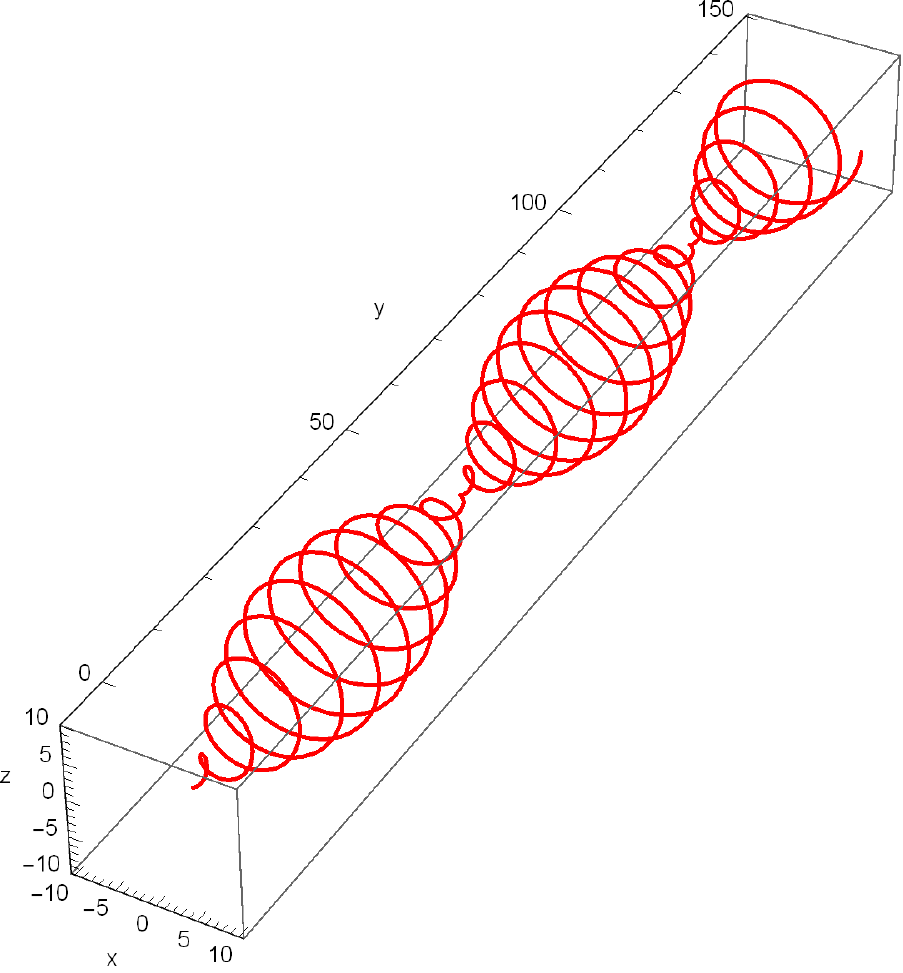}
				\captionof{figure}{Trajectory of a charged particle subjected to fields (\ref{eq: campo E do arquivo 2}) and (\ref{eq: campo B do arquivo 2}), with the parameters $ M=E_0=B_{y}=Q=1 $, $\omega=1.1$ and initial conditions $ x(0)=y(0)=z(0)=\dot{x}(0)=\dot{y}(0)=\dot{z}(0)=0 $, in a time interval [$ 0,50\pi $].}
				\label{fig: trajetoria arq 3.2}
			\end{minipage}
		\end{center}
	\end{figure}

	Defining the values of the constants as $ E_0=Q=M=B_y=1 $, we get the trajectories shown in the figures \ref{fig: trajetoria arq 3.3} and \ref{fig: trajetoria arq 3.2}.
	
	The solutions (\ref{eq: sol da equação de movimento arq 3 x}), (\ref{eq: sol da equação de movimento arq 3 y}) and (\ref{eq: sol da equação de movimento arq 3 z}), can be obtained in Mathematica from the commands
	\begin{lstlisting}[frame=single,mathescape=true,numbers=none]
		
		F = {Fx, Fy, Fz}; r = {x[t], y[t], z[t]}; EF = {Ex, Ey, Ez}; BF = {Bx, By, Bz}; v = D[r, t];
		
		F = Q*(EF + Cross[v, BF])
		
		eq1 = F - M*D[r, {t, 2}] == 0 // Thread; eq1 // ColumnForm
		
		initial = {x[0] == 0, x'[0] == 0, y[0] == 0, y'[0] == 0, z[0] == 0, z'[0] == 0};
		
		fields = Thread /@ {BF -> {0, By, 0}, EF -> {E0*Cos[$ \omega $*t],E0*Sin[$ \omega $*t], 0}} // Flatten
		
		eqs = Join[eq1, inicial] /. feilds
		
		DSolve[eqs, {x[t], y[t], z[t]}, t] // Flatten // FullSimplify
		
	\end{lstlisting}
	
	From the solutions obtained, we used to build the simulation illustrated in the figure \ref{fig: trajetoria arq 3.3} The codes
	
	\begin{lstlisting}[frame=single,mathescape=true,numbers=none]
		
		values={M-> 1,E0-> 1,By-> 1,Q-> 1}
		
		eqx[t_] := (E0 M Q (Cos[(By Q t)/M]-Cos[t $ \omega $]))/(-By^2 Q^2 + 	M^2 $ \omega $^2);
		eqy[t_] := (E0 Q (t $ \omega $ - Sin[t $ \omega $]))/(M $ \omega $^2);
		eqz[t_] := (E0 Q (-M $ \omega $ Sin[(By Q t)/M] + By Q Sin[t $ \omega $]))/(By^2 Q^2 $ \omega $ - M^2 $ \omega $^3);
		
		ParametricPlot3D[{eqx[t], eqy[t], eqz[t]} /. $\omega$ -> 0.99 /. valores, {t, 0, 20*Pi}, AxesLabel -> {"x", "y", "z"}, Mesh -> None, PlotRange -> All, PlotStyle -> {Red, Directive[EdgeForm[]], Thickness[0.005],		Opacity[1]}, ImageSize -> Medium]
		
	\end{lstlisting}
	from which we can obtain the trajectory of the figure \ref{fig: trajetoria arq 3.2} analogously, varying the value of $\omega$.

	\subsubsection{\label{subsubsec: 5}$\vec{E}$ oscillating and $\vec{B}$ uniform, in two directions $ (xy) $}
	
	Consider a charged particle with charge $ Q $ and mass $ M $, subjected to an oscillating electric field and a uniform magnetic field, both in two directions,
	\begin{align}
		\label{eq: campo E do arquivo 4}
		\vec{E}&=E_{0}\cos(\omega t)\hat{i}+E_{0}\sin(\omega t)\hat{j}\\
		\label{eq: campo B do arquivo 4}
		\vec{B}&=B_{x}\hat{i}+B_{y}\hat{j}.
	\end{align}

	Compared to the previous case, there has only been an addition of one component $ B_{x} $. However, this subtle change is enough to greatly complicate the solution of the equations of motion. Therefore, this is a problem that can be solved efficiently and much more practically by using Mathematica software than by performing the algebraic calculations by hand.
	
	For this system, the resulting equations of motion are given by
	\begin{subequations}
		\label{eq: equação de movimento arq 4}
		\begin{align}
			&Q\left[E_0 \cos(\omega t)-B_y\dot{z}\right]-M\ddot{x}=0\\
			&Q\left[E_0\sin(\omega t)+B_x\dot{z}\right]-M\ddot{y}=0\\
			&Q\left[B_y\dot{x}-B_x\dot{y}\right]-M\ddot{z}=0
		\end{align}
	\end{subequations}
	
	Although the system (\ref{eq: equação de movimento arq 4}) of coupled ODEs is very laborious to solve, there is a general analytical solution. Assuming that the initial conditions are $x(0)=y(0)=z(0)=\dot{x}(0)=\dot{y}(0)=\dot{z}(0)=0$, the general solution is obtained by means of the commands
	\begin{lstlisting}[frame=single,mathescape=true,numbers=none]
		
		F = {Fx, Fy, Fz}; r = {x[t], y[t], z[t]}; EF = {Ex, Ey, Ez}; BF = {Bx, By, Bz}; v = D[r, t];
		
		F = Q*(EF + Cross[v, BF])
		
		eq1 = F - M*D[r, {t, 2}] == 0 // Thread; eq1 // ColumnForm
		
		initial = {x[0] == 0, x'[0] == 0, y[0] == 0, y'[0] == 0, z[0] == 0, z'[0] == 0};
		
		feilds = Thread /@ {BF -> {B_x, By, 0}, EF -> {E0*Cos[$ \omega $*t],E0*Sin[$ \omega $*t], 0}} // Flatten
		
		eqs = Join[eq1, initial] /. feilds
		
		DSolve[eqs, {x[t], y[t], z[t]}, t] // Flatten // FullSimplify
		
	\end{lstlisting}
	Defining the constants $ A_{1},\ A_{2},\ A_{3},\ A_{4},\ A_{5},\ A_{6},\ k_{1}\ \text{e}\ k_{2} $ as
	\begin{align}
		A_{1}&=\frac{1}{M \omega ^2 \left[B_x^2+B_y^2\right] \left[Q^2 \left(B_x^2+B_y^2\right)-M^2 \omega^2\right]}\\
		A_{2}&=\frac{\sqrt{B_x^2+B_y^2}}{\sqrt{-\left(B_x^2+B_y^2\right)}}\\
		A_{3}&=Q^2\left(B_x^2+B_y^2\right)-M^2\omega^2\\
		A_{4}&=\frac{B_x^2 M^3\omega^3}{Q\sqrt{B_x^2+B_y}^2}\\
		A_{5}&=B_xB_yQ^2\left(B_x^2+B_y^2\right)\\
		A_{6}&=\sqrt{B_x^2+B_y^2}\\
		k_{1}&=-\frac{Q\sqrt{-(Bx^2 + By^2)}}{M}\\
		k_{2}&=\frac{Q\sqrt{B_x^2+B_y^2}}{M}
	\end{align}
	the general solution is given in the form
	
	\begin{widetext}	
	\begin{align}
		\label{eq: sol da equação de movimento arq 5 x}
	x(t)=&-A_{1}E_0Qe^{k_{1}t}\left[\cos\left(k_{2}t\right)-A_{2}\sin\left(k_{2}t\right)\right]\bigg[-B_xA_{3}(B_x+B_y\omega t)+B_y^2M^2\omega^2\cos\left(k_{2}t\right)\nonumber\\
	&+\left(B_x^2+B_y^2\right)(B_xQ-M\omega)(B_xQ+M\omega)\cos(\omega t)+A_{4}B_y\sin(k_{7}t)+A_{5}\sin(\omega t)\bigg]\\
		\label{eq: sol da equação de movimento arq 5 y}
	y(t)=&A_{1}E_0Qe^{k_{1}t}\left[\cos\left(k_{2}t\right)-A_{2}\sin\left(k_{2}t\right)\right]\bigg[B_yA_{3}(B_x+B_y\omega t)+B_xB_yM^2\omega^2\cos\left(k_{2}t\right)\nonumber\\
	&-\left(B_x^2+B_y^2\right)(B_yQ-M\omega)(B_yQ+M\omega)\sin(\omega t)+A_{4}B_x\sin(k_{7}t)-A_{5}\cos(\omega t)\bigg]\\
		\label{eq: sol da equação de movimento arq 5 z}
	z(t)=&A_{1}E_0Qe^{k_{1}t}\bigg\{\sqrt{B_x^2+B_y^2}\cos\left(k_{2}t\right)+\sqrt{-(B_x^2+B_y^2)}\sin\left(k_{2}t\right)\bigg\}\bigg\{-B_xA_{6}A_{3}-B_xA_{6}M^2\omega^2\cos\left(k_{2}t\right)\nonumber\\
	&+Q\left(B_x^2+B_y^2\right)\left[Q\sqrt{B_x^2+B_y^2}(B_x\cos(\omega t)+B_y\sin(\omega t))-B_yM\omega\sin\left(k_{2}t\right)\right]\bigg\}
	\end{align}
\end{widetext}
	
\begin{figure}[H]
			\centering
		\begin{minipage}{8cm}
			\centering
			\includegraphics[width=1\textwidth,height=5cm]{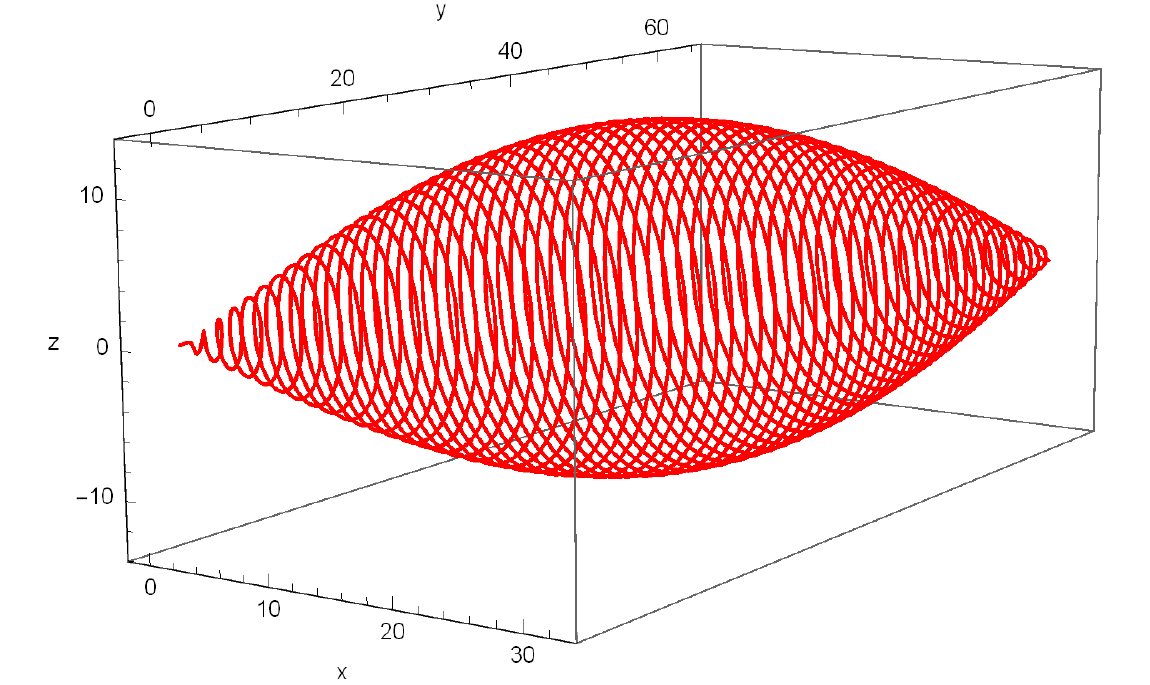}
			\captionof{figure}{Trajectory of a charged particle subjected to fields (\ref{eq: campo E do arquivo 4}) and (\ref{eq: campo B do arquivo 4}), with the parameters $ M=E_0=B_{x}=Q=1 $, $\omega=2.2$ and initial conditions $x(0)=y(0)=z(0)=\dot{x}(0)=\dot{y}(0)=\dot{z}(0)=0$, in a time interval [$ 0,55\pi $].}
			\label{fig: trajetoria arq 4.2}
		\end{minipage}
	\end{figure}

	With the solutions (\ref{eq: sol da equação de movimento arq 5 x}), (\ref{eq: sol da equação de movimento arq 5 y}) and (\ref{eq: sol da equação de movimento arq 5 z}) at hand, we can make computer simulations of possible trajectories performed by the charged particle, as shown in the figure \ref{fig: trajetoria arq 4.2}.
	
	For the figure to be obtained, it is necessary to define the value of the parameters of the equation,
	\begin{lstlisting}[frame=single,mathescape=true,numbers=none]
		values={M->1,E0->1,Bx->1,By->1,Q->1}
	\end{lstlisting}
	followed by the ParameticPlot3d command with input on the solutions obtained,
	\begin{lstlisting}[frame=single,mathescape=true,numbers=none]
		
		ParametricPlot3D[{eqx[t], eqy[t], eqz[t]} /. $\omega$ -> 2 /. values, {t, 0, 20*Pi}, AxesLabel -> {"x", "y", "z"}, Mesh -> None, PlotRange -> All, PlotStyle -> {Red, Directive[EdgeForm[]], Thickness[0.005],	Opacity[1]}, ImageSize -> Medium]		
	\end{lstlisting}

	\subsubsection{\label{subsubsec: 6}$\vec{E}$ oscillating in two directions $ (xy) $ and $\vec{B}$ uniform in three $ (xyz) $}
	
	Considering the case treated in the example \ref{subsubsec: 6}, an oscillating electric and uniform magnetic field in two directions, differing only in the constant magnetic field superposition on the $ z $ axis, i.e,
	\begin{align}
		\label{eq: campo E do arquivo 5}
		\vec{E}&=E_{0}\cos(\omega t)\hat{i}+E_{0}\sin(\omega t)\hat{j}\\
		\label{eq: campo B do arquivo 5}
		\vec{B}&=B_{x}\hat{i}+B_{y}\hat{j}+B_{z}\hat{k},
	\end{align}
	a força de Lorentz nos vela as equações de movimento
	\begin{subequations}
		\label{eq: equação de movimento arq 5}
		\begin{align}
			&Q\left[E_0 \cos(\omega t)+B_z\dot{y}-B_y\dot{z}\right]-M\ddot{x}=0\\
			&Q\left[E_0\sin(\omega t)-B_z\dot{x}+B_x\dot{z}\right]-M\ddot{y}=0\\
			&Q\left[B_y\dot{x}-B_x\dot{y}\right]-M\ddot{z}=0
		\end{align}
	\end{subequations}
	
	Mathematica finds the general solution of the system (\ref{eq: equação de movimento arq 5}) with the codes 
	\begin{lstlisting}[frame=single,mathescape=true,numbers=none]
		
		F = {Fx, Fy, Fz}; r = {x[t], y[t], z[t]}; EF = {Ex, Ey, Ez}; BF = {Bx, By, Bz}; v = D[r, t];
		
		F = Q*(EF + Cross[v, BF])
		
		eq1 = F - M*D[r, {t, 2}] == 0 // Thread; eq1 // ColumnForm
		
		initial = {x[0] == 0, x'[0] == 0, y[0] == 0, y'[0] == 0, z[0] == 0, z'[0] == 0};
		
		feilds = Thread /@ {BF -> {B_x, By, Bz}, EF -> {E0*Cos[$ \omega $*t],E0*Sin[$ \omega $*t], 0}} // Flatten
		
		eqs = Join[eq1, initial] /. feilds
		
		DSolve[eqs, {x[t], y[t], z[t]}, t] // Flatten // FullSimplify
		
	\end{lstlisting}
	\begin{widetext}
	where, we assume that the initial conditions are $x(0)=y(0)=z(0)=\dot{x}(0)=\dot{y}(0)=\dot{z}(0)=0$.
	
	Defining the constants
	\begin{align}
	&A_{1}=\frac{1}{2MQ\omega^2\left[B_x^2+B_y^2+B_z^2\right]^{5/2}\left[Q^2\left(B_x^2+B_y^2+B_z^2\right)-M^2\omega^2\right]}\\
	&A_{2}=\sqrt{B_x^2+B_y^2+B_z^2}\left[Q^2\left(B_x^2+B_y^2+B_z^2\right)-M^2\omega^2\right]\\
	&A_{3}=B_x^2+B_y^2+B_z^2\\
	&A_{4}=+2M^2\omega^2\sqrt{B_x^2+B_y^2+B_z^2}\\
	&A_{5}=M^2\omega^2(-A_{3}E_zQ^2-B_xE_0MQ\omega+E_zM^2\omega^2)\\
	&A_{6}=2B_yM^2\omega^2\sqrt{B_x^2+B_y^2+B_z^2}\left[A_{3}(B_xE_0+B_zE_z)Q^2-B_zE_zM^2\omega^2\right]\\
	&A_{7}=M^2\omega^2 \left[(B_x^2 + B_y^2 + B_z^2)(-B_zE_0+B_xE_z)Q^2+(B_x^2+
	B_z^2)E_0MQ\omega-B_xE_z M^2\omega^2\right]\\
	&A_{8}=A_{3}^{3/2}E_0Q^2\left[B_y^2Q^2+M\omega(B_zQ-M\omega)\right]\\
	&A_{9}=2M^2\omega^2\left\{A_{3}Q^2\left[-B_xB_zE_0+(A_{3}-B_z^2)E_z\right]+B_xA_{3}E_0MQ\omega-(A_{3}-B_z^2)E_zM^2\omega^2\right\}\\
	&A_{10}=\frac{1}{\left[2A_{3}^2\omega^2(A_{3}MQ^3-M^3Q\omega^2)\right]}\\
	&k_{1}=\sfrac{\sqrt{A_{3}} Q}{M}\\
	&k_{2}=\sfrac{Q\sqrt{B_x^2+B_y^2+B_z^2}}{m}
	\end{align}
	we can write the solution in the form
	\begin{align}
	\label{eq:sol7x}
	x(t)=&A_{1}\biggl\{2A_{2}\bigg[2E_{0}Q^{2}B_x^{2}+2A_{3}E_0Q\omega (B_x B_y Q t+B_z M)+E_z \omega^2 \left(B_x B_zQ^2A_{3} t^2-2B_y M Q A_{3}t -2 B_x B_zM^2\right)\bigg]\nonumber\\
	&+A_{4}\cos(k_{1}t)\bigg[A_{3}Q^2\left(B_xB_zE_z-E_0B_y^2-E_0B_z^2\right)+A_3B_zE_0MQ\omega-B_xB_zE_zM^2 \omega^2\bigg]-2A_{3}^{5/2}E_0 B_x^2Q^4 \nonumber\\
	&-2A_{3}^{5/2}E_0 Q^2M\omega (B_z Q -M\omega)\cos(\omega t)-2B_yA_{3}\left[A_{5}\sin(k_{1}t)+B_xE_0A_{3}^{2/3} Q^4 \sin(\omega t)\right]\biggr\}\\
	\label{eq:sol7y}
	y(t)=&A_{1}\biggl\{A_{2}\bigg[2B_xB_yA_{3}E_0Q^2+2A_{3}E_0Q^{2}B_y^{2}\omega t+E_z\omega^2\left(-2B_yB_zM^2+2B_xA_{3}MQt+B_yB_zA_{3}Q^2t^2\right)\bigg]\nonumber\\
	&+A_{6}\cos(k_{1}t)-2A_{3}\left[B_{x}B_{y}A^{3/2}_{3}E_{0}Q^{4}\cos(\omega t)+A_{7}\right]\sin(k_{1}t)+A_{8}\sin(\omega t)\biggr\}\\
	\label{eq:sol7z}
	z(t)=&A_{10}\biggl\{\bigg[A_{3}Q^2-M^2\omega^2\bigg]\bigg[2B_xB_zA_{3}E_0Q^2-2A_{3}E_0Q(B_xM-B_yB_zQt)\omega+2E_z(B_x^2+B_y^2)M^2+E_zB_z^2A_{3}Q^2t^2\omega^2\bigg]\nonumber\\
	&-A_{9}\cos(k_{2}t)-2E_0QA_{3}^{1/2}\bigg[B_xA_{3}^{3/2}Q^2(B_zQ-M\omega)\cos(\omega t)+B_y M^2\omega^2(A_{3}Q-B_zM\omega)\sin(k_{2}t)\nonumber\\
	&+B_yA_{3}^{3/2}Q^2(B_zQ-M\omega)\sin(\omega t)\bigg]\biggr\}
	\end{align}	
	\end{widetext}

	In order to get the simulation, we will set the parameters $ M=E_0=B_{x}=B_{y}=B_{z}=Q=1 $. This can be done in Mathematica as follows,
	\begin{lstlisting}[frame=single,mathescape=true,numbers=none]
		
		values={M->1,E0->1,Bx->1,By->1,Q->1}
	\end{lstlisting}

	Using ParametricPlot3d, we can call the solutions (\ref{eq:sol7x}), (\ref{eq:sol7y}) and (\ref{eq:sol7z}), to generate the trajectory of the particle, by means of the commands\vspace{12pt}
	\begin{lstlisting}[frame=single,mathescape=true,numbers=none]
		
		ParametricPlot3D[{eqx[t], eqy[t], eqz[t]} /. $\omega$ -> 2 /. values, {t, 0, 20*Pi}, AxesLabel -> {"x", "y", "z"}, Mesh -> None, PlotRange -> All, PlotStyle -> {Red, Directive[EdgeForm[]], Thickness[0.005],		Opacity[1]}, ImageSize -> Medium]
	\end{lstlisting}
	
\begin{figure}[H]
		\begin{center}
			\includegraphics[width=0.45\textwidth]{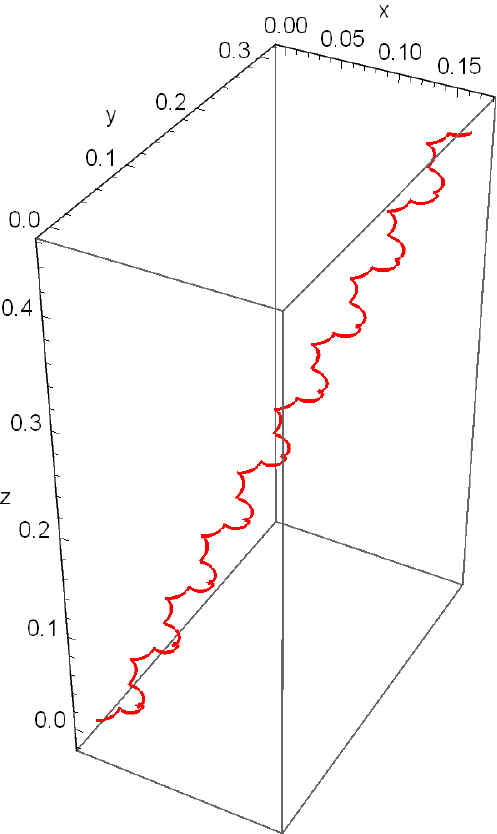}
			\begin{minipage}{8cm}
			\centering
			\caption{Trajectory of a charged particle subjected to fields (\ref{eq: campo E do arquivo 5}) and (\ref{eq: campo B do arquivo 5}), with the parameters $ M=E_0=B_{x}=B_{y}=B_{z}=Q=1 $, $\omega=15$ and initial conditions $x(0)=y(0)=z(0)=\dot{x}(0)=\dot{y}(0)=\dot{z}(0)=0$, in a time interval [$ 0,5\pi $].}
			\label{fig: trajetoria arq 6.1}
			\end{minipage}
		\end{center}
	\end{figure}

	The result is the trajectory illustrated in the figure \ref{fig: trajetoria arq 6.1}.
	
	\subsubsection{\label{subsubsec: 7}Simulation for an oscillating electric field}

We can simulate an oscillating electric field of the case in the figure \ref{fig: trajetoria arq 2.4}, where $M=E_0=B_x=Q=1$ and $\omega=0.51$. Through the commands 
\begin{lstlisting}[frame=single,mathescape=true,numbers=none]
x[t_] := -((E0 q (-1 + Cos[t $\omega$]))/(m $\omega$^2)); 
y[t_] := (E0 m (-m $\omega$ Sin[(Bx q t)/m] + Bx q Sin[t $\omega$]))/( Bx^3 q^2 - Bx m^2 $\omega$^2); 
z[t_] := (E0(-Bx^2q^2+m^2$\omega$^2-m^2$\omega$^2Cos[(Bx q t)/m]+
    Bx^2 q^2 Cos[t$\omega$]))/(Bx^3 q^2 $\omega$-Bx m^2 $\omega$^3);
m = 1; q = 1; E0 = 1; Bx = 1;
tRange=Range[0.001,100*Pi,100*Pi/500];$\omega$=0.51;
For[k = 1, k <= Length[tRange], k++, tk=tRange[[k]]; xk=x[tk];yk=y[tk];zk=z[tk]; 
 movieVector[k]=Show[Graphics3D[{AbsolutePointSize[15], Green, Point[{xk, yk, zk}]},
     PlotRange -> {{-0.2, 8.2}, {-2.2, 2.2}, {-5.7, 1.7}}], 
   ParametricPlot3D[{x[t], y[t], z[t]}, {t, 0, tk}, PlotStyle -> Red, 
    PlotRange -> {{-0.2, 8.2}, {-2.2, 2.2}, {-5.7, 1.7}}], 
   AxesLabel -> {"x", "y", "z"}, 
   PlotLabel -> StringJoin["t=", ToString[N[tk]]]]];
   M = Table[movieVector[k], {k, 1, 500}];  
   fileName = "FilmeMathematica.avi"; Export[fileName, M];
   Directory[]	
\end{lstlisting}
	
The resulting simulation can be seen in the following video https://youtu.be/nlUoLaV4ECc.

	\section{Conclusion}
	\label{sec:4}
	
In this paper we propose the use of Mathematica software to solve ordinary differential equations, and to represent the trajectories of charged particles subjected to oscillating electric and constant magnetic fields. 
\par 
We begin by showing the use of Mathematica software in solving the exact differential equation of the simple harmonic oscillator, and the graphical representation of this solution. Then we show a numerical solution of the simple pendulum, also showing the graphical representation of that solution and its phase space. 
\par 
Finally, we show how to use Mathematica software to construct the differential equations of Newton's second law of motion, which is the Lorentz force, defining position vector, velocity, electric and magnetic fields, and thus solve the equations of motion, calling the solutions to represent them graphically, as trajectories of the charged particle. 
\par 
We end by showing how to build a video simulation of the charged particle trajectory representation. 
\par 
We believe that this article serves as an introduction to symbolic algebraic computation, for use in topics that are more difficult to obtain non-analytic solutions, in several areas of Physics and Mathematics. The generalization of its use to solve partial differential equations, and nonlinear equations is possible.

	
	\section{Acknowledgments}
	The authors thank CNPq for the PIBIC scholarship and partial financing of this work.
	\label{agrad}


\end{document}